%% file: main.tex
\pdfoutput=1 
\documentclass[runningheads]{llncs}
\usepackage{graphicx}
\usepackage[hidelinks]{hyperref}
\usepackage[space]{cite}
\usepackage{csquotes}

\usepackage{booktabs} 

\begin{document}
\title{``I Don't Know Too Much About It'': On the Security Mindsets of Computer Science Students}

\titlerunning{On the Security Mindsets of Computer Science Students}
%
\author{Mohammad Tahaei \and
Adam Jenkins \and
Kami Vaniea \and Maria Wolters}
\authorrunning{M. Tahaei et al.}
%

\institute{School of Informatics, University of Edinburgh \quad \\
\email{\{mohammad.tahaei, adam.jenkins, kami.vaniea, maria.wolters\}@ed.ac.uk}}

\maketitle              
\begin{abstract}
The security attitudes and approaches of software developers have a large impact on the software they produce, yet we know very little about how and when these views are constructed. This paper investigates the security and privacy (S\&P) perceptions, experiences, and practices of current Computer Science students at the graduate and undergraduate level using semi-structured interviews. We find that the attitudes of students already match many of those that have been observed in professional level developers. Students have a range of hacker and attack mindsets, lack of experience with security APIs, a mixed view of who is in charge of S\&P in the software life cycle, and a tendency to trust other peoples' code as a convenient approach to rapidly build software. We discuss the impact of our results on both curriculum development and support for professional developers.

\end{abstract}

\section{Introduction}
\label{introduction}
Software developers can impact millions of lives with seemingly small security decisions that have a large impact on the people using the technologies. One example is the case of the dating site Ashley Madison, where a strong cryptographic algorithm was used to store passwords but was implemented incorrectly~\cite{mansfield2015ashley}. 

Even for apps where security is not a primary feature, it is a requirement needed for stability and safety of operation. 
Therefore, software developers need to be keenly aware of the security implications of their design decisions. Ideally, they should have strong support from their tools to avoid security and privacy issues in their resulting code.

Basic tools such as cryptographic libraries (OpenSSL) and federated authentication (OAuth) exist partially to assist developers in integrating common security needs into their projects without needing to know all the complex details. There are also efforts to help raise awareness of common coding and design issues such as the IEEE top ten security flaws~\cite{arce2014avoiding,acar2017developers}. 

Yet, security remains a pervasive problem in deployed code. In 2013 alone, $88\%$ of apps (out of 11,748) analysed on Google Play had at least one mistake in how the developer used a cryptographic API~\cite{egele2013empirical}. Code that they write goes into security-critical applications such as banking software~\cite{georgiev2012most} as well as software with less obvious security implications such as Internet connected kettles~\cite{munro2015kettles}.

Non-usable APIs are a key point of failure for most developers~\cite{wurster2009developer,indela2016toward,green2016developers,acar2016dev,pieczul2017developer}. Providing manuals is not enough. A usability evaluation of programming security in Android found that developers created code with security errors even when they were provided with official documentation~\cite{acar2016impact}. Perhaps more importantly, developer understanding of security is also problematic. Interviews with professional developers show a range of concern about security and privacy knowledge~\cite{balebako2014improving}. The situation is exacerbated when developers make non-obvious errors when implementing security which results in believing that code is secure when it is actually not secure~\cite{acar2017comparing}.

One potential opportunity for changing developers' security attitudes and practices is during their training. In this work, we investigate the security and privacy (S\&P) mindsets of a group of twenty graduate and undergraduate computer science (CS) students on a variety of career trajectories, and with a range of exposure to formal security training. 
Our research questions are: 
\begin{itemize}
    \item What are students' comprehension of S\&P related concepts?
    \item To what extent do students consider S\&P while coding applications, and how do they implement it?
\end{itemize}

Within the context of developer-centred security, our study highlights the extent to which students already have similar mindsets and practices as have been found in professional developers, suggesting that these may form and consolidate early. We conclude that, while early educational intervention would be ideal, we also need to provide developers with usable tools, such as APIs, and easily accessible training, which can be used both by trainees and professionals.

\section{Related work}
\label{sec:relatedWorks}
Creating secure software correctly is quite challenging even for professional developers, often resulting in unintended security vulnerabilities~\cite{wurster2009developer,green2016developers,acar2016dev}.
The OWASP organisation publishes the top ten most critical web application security risks every few years. A review of their last three reports covering seven years terrifyingly that the most common issues are quite stable~\cite{owasp2017top}, with common and highly damaging vulnerabilities such as code injection and broken authentication continuously remaining in the top ten.

Arce et al. observed that many of the OWASP vulnerabilities represent unintentional errors or mistakes rather than planned actions and therefore are minimally helpful to someone trying to design a secure system~\cite{arce2014avoiding}. Instead they propose a set of top ten \emph{security design flaws}, that is security issues that are a planned element of the software. Their list is much higher-level and contains issues such as ``earn or give, but never assume, trust''~\cite[p.~9]{arce2014avoiding}. 

The problem of code vulnerabilities in live software is further exacerbated by the steady reduction of the barriers to entry for new software creators. While generally a good thing, the `anyone can code' movement has also led to an increase in the number of software creators with minimal formal training in software development and even less training in security. Unsurprisingly, this group also has difficulty creating secure software~\cite{pieczul2017developer,oltrogge2018rise}. 

Neither of these groups is, or should be, expected to be security experts, but the decisions they make can still have serious security impacts. In an effort to better support these software creators, several tools and libraries have been proposed such as OpenSSL, PyCrypto, and cryptography.io which encapsulate many of the security decisions, theoretically making development easier.

Unfortunately, many of these tools still suffer from usability issues, such as confusing API designs~\cite{georgiev2012most,fahl2013rethinking,egele2013empirical,lazar2014does,acar2017comparing,iacono2017and,ukrop2018johnny} or poorly designed documentation~\cite{acar2016impact,nadi2016jumping}. Official documentations are often not easy to use, hence developers prefer online resources which may not offer valid and secure solutions. While Stack Overflow, for example, helps with getting code working quickly, the suggested solutions may also result in less secure code~\cite{acar2016impact,fischer2017stack}.

Security is also challenging for developers because it causes no obvious visual effect, making it difficult to identify when an unintended state has occurred~\cite{fahl2012eve,fahl2013rethinking}. 
A common example of invisible security effects is SSL/TLS. When used incorrectly, a connection is still formed, but that connection might not be encrypted, or it might be encrypted, but without certificate validation. This results in a vulnerability to man-in-the-middle (MITM) attacks during connection setup. 
Fahl et al. observed how challenging this can be for developers to spot. One of their developers even used Wireshark to `test' the SSL/TLS connection and, because the data was garbled looking, incorrectly concluded things were working even though no certificate checking was happening~\cite{fahl2013rethinking}.

Georgiev et al. similarly conducted an analysis of SSL certificate validation in multiple non-browser platforms and showed that many applications on the market are open to a MITM attack where data can be read and modified in transit because developers accidentally or intentionally configure their code to not validate the certificate source~\cite{georgiev2012most}. 
Such problems arise when developers are expected to understand the implications of the different settings of SSL, which is exacerbated by APIs that do not offer a helpful level of abstraction~\cite{iacono2017and}.

Security is also not a well-established requirement in the software development workflow. Without a dedicated developer in charge, security becomes a hot potato which is passed between groups because no one wants to deal with it~\cite{balebako2014improving,oliveira2014s,weir2016improve,poller2017can}. In interviews with security experts, Thomas et al. found that security auditing is seen as a separate task from software development. While security auditing is performed by the rare breed that are security experts, it is then the developer's job to fix the security issues found~\cite{thomas2018security}.

Many future software developers were once Computer Science (CS) students. A survey by Stack Overflow in 2019 showed that $62.4\%$ (75,614 responses) of developers have a degree in CS, computer engineering, or software engineering~\cite{stackoverflow2019dev}. Given the importance of this group, many researchers study them to either address gaps between academia and industry~\cite{radermacher2013gaps,radermacher2014investigating,cambazoglu2013computer,sudol2010analyzing,jones2017should} or to suggest educational tools to improve their skill and abilities~\cite{whitney2015embedding,nielson2016playground,tabassum2018evaluating}.
Research shows that CS students often work under misconceptions which can lead to bad practice. For example, when it comes to software engineering processes and teamwork~\cite{sudol2010analyzing}, many think that working alone is a quicker way of working on a software project, which goes against established industry best practice. Here we study the S\&P mindsets of CS students with a view to identifying what they know and think about S\&P, and what misconceptions exist.

\section{Methodology}
\label{sec:methodology}
We used semi-structured interviews to explore how a range of students from undergraduate to PhD think about S\&P. The semi-structured approach allowed us to probe students' S\&P mindsets in detail and investigate how they relate to their own practices as developers. 

\subsection{Interview design}
After informed consent, we explicitly invited participants to talk as much as they wanted on the various topics discussed. The interview began with an open question on academic and professional background and general questions about coding and software development experience. Questions about demographics were asked at the end of the interview in order to minimise stereotype threat. The full interview script is included in the~\hyperref[interview-guideline]{Appendix}.

We began the S\&P discussion by asking participants to consider creating ``a new group discussion app for in-class discussions.'' They were then asked to free-list the app's features on paper and after they finished they were asked to circle those that were S\&P related. 

Next, we examined participants' understandings around threats and hackers. We started by asking participants about the hypothetical app: ``Who is most likely to try and attack this system? What are they likely going to try and do?''
We then moved on to talk about hackers, because work on security folk models has found them to be an important part of how people think about security~\cite{wash2010folk}. We elicited participants' definitions of the term hacker, and their views on hackers' intentions, goals, and background. 

We then moved on to considering who was responsible for S\&P in software development practice. The discussion was grounded in participants' own experience of writing software, in particular problems with (security) APIs. 

Finally, we asked participants about personal security and privacy practices. First, participants were asked to list the words and concepts they associated with `computer security' on paper. We followed up with questions about good security practices, and their own security practices. 
 
Since prior negative experiences can impact future choices~\cite{vaniea_betrayed_2014}, we also asked about prior experiences with compromise, prompting them with examples such as ``getting a virus on your computer, losing your password, having an email sent from your account, or loss of data about you'' if needed. We explored how the experience was resolved, and what participants learned. 

\subsection{Recruitment}
We recruited participants through mailing lists associated with a large Russell Group University in the United Kingdom, Facebook groups, and word of mouth. 
Advertisements asked for Computer Science students (BSc, MSc and PhD) to participate in an interview about opinions and attitudes around software development, particularly around the handling of requirements prioritisation. All advertisements avoided S\&P related words to limit self-selection and priming. 

\subsection{Participants}
Our sample, shown in Table~\ref{tab:participants-demo}, includes twenty students (6 BSc, 11 MSc, and 3 PhD students), participants who previously took a computer security course at any University are indicated with `PS' instead of `P'. The sample contains five female, and fifteen 
male students with an average age of 24 years old (range: $20-37$, std: $3.8$, median: $23$). They 
come from various countries and have diverse CS-related educational backgrounds. Interviews were conducted in English. Our sample reflects both the diversity seen in the tech industry~\cite{google2018diversity, Azhar2019Securing}, and
the culturally diverse classrooms found in many computer science departments.

The interviews were advertised to be 60 to 90 minutes long with a compensation 
of \pounds10 in cash. In practice, interviews took an average of 68 minutes (range: $41-108$, std: $18.4$, median: $65.5$) and were completed in July 2018. All interviews were audio recorded with participant consent. The study was conducted in accordance with the Ethics procedures of the School where the students were recruited (cert ID 2870).

We interviewed students over the summer. This meant that the Masters students were in their dissertation phase, and had completed the course work part of their 12-month degree. PhD students in the UK have typically completed a Masters before starting a PhD and are not necessarily required to take courses, pass a qualifying exam, or be a Teaching Assistant, though many choose to take additional courses and tutor. Therefore, beyond teaching and thesis work, PhD students are unlikely to be impacted by security courses taught at the University.

\begin{table}[ht]
  \caption{Interview study demographics. P = participant without computer security background; PS = participant who self-describes as having taken a computer security course in the past. }
  \label{tab:participants-demo}
  \centering
  \begin{tabular}{l c c c c}
    \toprule
    Participant & Gender & Nationality & Age & Expected Degree \\
    \midrule
    PS01 & M & EU & 29 & PhD \\ 
    P02 & M & EU & 28 & MSc \\
    PS03 & F & Asia & 22 & MSc \\
    PS04 & M & Asia & 24 & MSc \\
    PS05 & M & Asia & 25 & PhD \\
    P06 & F & Asia & 23 & MSc \\
    P07 & M & Asia & 22 & BSc \\ 
    PS08 & M & UK & 21 & MSc \\
    PS09 & M & Asia & 25 & MSc \\
    P10 & M & Asia & 21 & BSc \\
    P11 & M & EU & 22 & BSc \\ 
    PS12 & M & Asia & 23 & MSc \\
    PS13 & M & EU & 21 & BSc \\
    P14 & M & EU & 20 & BSc \\ 
    PS15 & M & EU & 25 & PhD \\
    PS16 & M & Asia & 37 & MSc \\
    P17 & F & EU & 25 & BSc \\
    P18 & F & Asia & 23 & MSc \\
    P19 & M & UK & 24 & MSc \\
    P20 & F & Asia & 20 & MSc \\
  \bottomrule
\end{tabular}
\end{table}

\subsection{Pilot}
We conducted seven pilot interviews with Masters and PhD students, six of which were associated with our research lab but unfamiliar with the work. These 
interviews were used to iteratively refine the interview script as well as adjust the number and content of questions to keep interviews at about 60 minutes. 
The pilot contained some students with no security background to help ensure the phrasing of security questions was clear. 
Feedback was also sought about the structure, clarity, and accuracy of the interview schedule. Pilot interviewees and interviews were not used in our final analysis.

\subsection{Interview analysis}
Interview analysis focused on uncovering students' mindsets of S\&P as they relate to the software development process. Relevant themes were extracted using a three stage process. First, two researchers listened to the full audio of four interviews which had been selected by the interviewer to cover a wide range of participants, identified relevant parts for more detailed analysis and transcription, and outlined an initial topic guide for coding~\cite{miles1994qualitative,saldana2015coding}. Audio was used because it provides a richer record of the original interview than a standard transcript. In the second stage, the researchers performed open coding of the transcripts based on the topic guide~\cite{miles1994qualitative,saldana2015coding}. 

In the third stage, the open codes were analysed using an affinity diagram~\cite{lazar2017research} to yield a set of seven themes, which are discussed in the Results Section~\ref{sec:results} below. While some authors suggest reporting how many participants mention each theme~\cite{lazar2017research}, we chose to follow standard qualitative research reporting practice and focus on describing and contextualising our themes~\cite{wash2010folk,renaud2014doesn}.

\section{Results}
\label{sec:results}
All participants all had some form of prior programming experience ranging from classroom projects, internships, and prior employment in industry. Since our participants included a large number of Masters students, they also had classroom experience from prior universities, with several expressing that they had worked in industry either as interns or full time before coming back for a Masters or PhD. Half had taken a computer security course at some point in their education. We did not ask about the details of these courses.

\subsection{`Computer security' word association results} 
\label{subsec:words}
Mid-way through the interview participants were asked to free-list words associated with `computer security'. The words were grouped into topics by the lead researcher with a bottom-up approach. A second researcher then reviewed the groupings and disagreements were resolved through discussion.
Table~\ref{tab:words} shows the resulting eleven topics.

Participants' understanding of the term `computer security' was broad, with participants who wrote words providing an average of $9.6$ words (range: $2-19$, std: $4.2$). 
Listed words included standard security topics such as encryption, attacks, and system security which are readily found in most security text books. Participants also listed company names that are either associated with security (Norton) or that had been discussed recently in the news in relation to security (Facebook~\cite{vox2018facebook,icoDataAnalytics}). Two participants (P02 \& P20) were not able to list any words, suggesting uncertainty with the term. \textquote[P02]{It is all very flimsy}, \textquote[P20]{To be honest I do not know too much about it}.

Of the participants who provided words, participants listed words from an average of $4.2$ topics (range: $1-7$, std: $1.8$).
The topics cover a wide range, but each individual participant had less range, with at most seven topics mentioned by one participant. Most notable is the lack of a single common topic amongst participants. For example, the most common word `privacy' was mentioned by only 40\% of participants. Common security topics such as passwords, authentication, and encryption also appeared. Some of these topics are similar to what professional developers associate with security, for example, encryption, user control, and user access~\cite{hadar2018privacy}.

\begin{table}
  \caption{Topics mentioned during free-listing, number of words participants listed associated with that topic, number of unique participants listing at least one word associated with the topic, and a set of sample words representing the range.}
  \label{tab:words}
  \resizebox{\columnwidth}{!}{%
  \centering 
  \def\arraystretch{1.2}
  \begin{tabular}{l c c p{8.5cm}}
    \toprule
    Topic & \#Words & \#Participants & Example words \\
    \midrule
    Encryption & 28 & 11 & 
        End-to-end, hash, RSA, public/private key, SSL, symmetric. \\
    Authentication & 28 & 9 & 
        Passwords, permissions, 2FA, tokens, access controls, emails. \\
    Privacy & 27 & 10 & 
        Anonymity, right to be forgotten, visibility, cookies. \\
    Attacks & 25 & 8 &
        Reconnaissance, phishing, buffer overflows, DoS, MITM. \\
    System security & 13 & 5 & 
        Protocols, database, Unix, system calls, TCP/IPs. \\
    Social & 13 & 7 & 
        Regulations, roles, responsibilities, public knowledge. \\
    Finance & 8 & 4 & 
        PayPal, Apple Pay, Bitcoin, online payments. \\
    Defending & 7 & 5 &
        Anti-virus/malware, penetration testing, logging, bounties. \\
    Security holes & 5 & 4 & 
        Failures, physical access, loopholes. \\
    Companies & 5 & 3 & 
        Facebook, Google, Norton, Red Hat. \\
    Trade offs & 4 & 3 & 
        Usable security, features vs security, easy to use UX. \\
  \bottomrule
\end{tabular}
}
\end{table}

\subsection{Interview themes}

\subsubsection{Security mindsets.}
Participants varied substantially in their understanding of S\&P. While some participants had a strong up-front understanding of security which varied minimally during the interview, others had clearly not thought much about the topic before resulting in them re-thinking their opinions mid-interview. This is to be somewhat expected as many people have not previously devoted extensive time to assessing their own understanding of the topic~\cite{asgharpour2007mental}. 

This theme provides rich additional context to the initial topics identified through free association. Those with a more sophisticated understanding of S\&P tended to use more definitive language, had more stable descriptions of attacker motivations, and were more likely to be sure that their statements were accurate, and to describe less intuitive or extreme scenarios. For example, PS15, a crpytography PhD student, explains that \textquote[PS15]{in crypto, we assume that the attacker is any code, literally any Turing machine}. 

Those with an initially less sophisticated understanding of S\&P showed signs of forming their opinions as the interview progressed. Often, this would involve contradictions in thoughts as they finally reached a definition for themselves. This was most notable for the hacking theme. Participants with less developed models exhibited less self-assurance around motivations, or definitions of attack scenarios. \textquote[P17]{I think [HTTPS] is standard by now, don't they? The more encryption the better? [...] Like exchange of data that's not encrypted at all. I don't think that's happening anymore. I'm not sure but I don't think it is}.

Similar to non-tech savvy users~\cite{wash2010folk,zou2018equifax}, some of our participants think they are not a target for attackers. \textquote[P11]{We are just average people. It is ok to have small security measures}, \textquote[P20]{I am also very boring computer user. I just do my courses and I watch movie on Netflix. So I don't really do anything that could put me in front of a virus}. Conversely, some participants had high awareness of potential attacks, though they still did not perceive themselves as at risk. \textquote[PS13]{I am running a server at home, which has an SSH access available. There you can see a lot of stuff going on, there are just bots or so whatever trying to get into. That is even a bit scary if you see that happen all the time, but I think my pass has been strong enough to keep them out}.

Participants clearly evidenced their own internal struggle over what S\&P actually was and when it was or was not needed, which might partially explain its lack of inclusion in initial requirements. \textquote[P06]{[My address] is not so important, because every website is required. Maybe because I live in a dormitory, if it is in my home that is different}.

While participants understood that private data should be protected, they struggled with what `private data' actually meant. Even when talking about S\&P in their private lives, participants had mixed opinions about how problematic it was for data like bank transactions to be leaked. \textquote[PS16]{So the data [leak] was about the full info about 
the bank accounts, the transactions, in and out, the current amount in it. For me it was normal [...] to have these transactions. But for some people it was an issue, because they receive money from hidden source, so it was an issue for them}. 

\subsubsection{Who are hackers and what do they want?}
Some participants' definitions of hackers were well articulated. 
\textquote[PS15]{Really theoretical let's say, the adversary we say in crypto is literally anyone that has a computer and some access to your systems}. Other participants had a more general understanding. \textquote[P02]{The images that you have in your head are from Hollywood. Super smart kids sitting in the corner of a room then CIA calls upon them to solve a problem}.

We found a wide range of imagined intentions for hacking, such as financial, personal, political, and just for fun. All four types of previously observed models of hackers from Wash's work~\cite{wash2010folk} were mentioned by our participants: 

Graffiti, which is a mischief causing attacker with technical background: \textquote[PS12]{Want to try what they learn from the class. They may write some code to hack some system of the school to show their ability}; Burglar, who commits crimes using computers mostly with financial motivations: \textquote[PS01]{There is nothing but personal interest. Personal gain. Personal satisfaction. And of course they are who just do it for financial gain. Stealing identities, pictures, personal info. Just to sell it afterwards, to like black market}; Big fish, who looks only for high valued targets: \textquote[PS13]{Political incentive that certain countries fund a lot of hacking and cracking to gain power depending how important or how famous you are there might be people who want to get access to your account}; and Contractor, a Graffiti hacker with financial/criminal motivations: \textquote[PS05]{Trained people who are trained to do this kind of stuff. Either by some governments to hack other governments. Or to break the encryption or security mechanism}.

\subsubsection{The role of security when planning software.}
When participants were asked about what features they would consider in an in-class discussion app, they commonly mentioned functional requirements including task management, calendar, question/answering, recording classes, and assignment management. 
Many of these features currently exist in course management software with which the students are familiar, such as Blackboard LEARN and Piazza.  

Only four participants (PS08, PS15, PS16, and P19) mentioned S\&P in their initial design and feature list, a somewhat small number since ten of our participants had previously taken a security course. Only two of the participants proactively brought up privacy issues. \textquote[P19]{First thing that comes to my mind is privacy. Definitely in 
terms of features. Presumably, the School will wish to host it locally rather than to have some sort of central cloud back service}, while PS08 noted the connection between privacy and ethics: \textquote[PS08]{There is some ethical questions involved in the area of student privacy}. 

Security of the data was also a concern, particularly in terms of information leaks. \textquote[PS16]{I will make sure of the safety and security because [no one] wants to use the tool if he feel he is vulnerable his info may leak to any unwanted person}. PS15 was also able to pull on prior experience and identify specific attacks and solutions that needed to be addressed: \textquote[PS15]{For sure I put HTTPS and TLS around it. So that would be safe. Because still, I would leave a lot of surface for attacks, because the big applications have more surface for attacks.... All those places where there is user input we basically talk about security, and we have to remember SQL injection and stuff like that}.

Some participants turned to more authoritative sources such as laws, regulations, and public policies as a guide for what should and had to be built into the system. \textquote[P02]{You have designed an app I guess you also think about security. But you also think about engagement. Does a certain security feature if it is option not legally required, how does it sort of effecting the engagement}. Some mentioned the EU \textit{General Data Protection Regulation} (GDPR, enforcement date: 25 May 2018)~\cite{gdpr2018eu}, either as a convenience tool for end-users or from a regulation perspective for companies. \textquote[P11]{Do we have to be GDPR compliance? Probably, I'm guessing} was mentioned by a participant when answering a question about what S\&P features his hypothetical classroom app might need.

\subsubsection{Requirements and responsibilities: playing hot potato.}
Several participants recognised security as an \textit{explicit requirement}. They consider the developers' job to be transforming requirements into code. Therefore if security is an explicit requirement, then they have to take it into account during design and development. \textquote[PS01]{So as a software engineer, if I am 
already given a certain requirement, I should not care about anything else outside the specs. You are employed as software engineer, you just write your. You are given a list, you just have to code it. Right? Unless you can do that. You are still doing your job}.

On the other hand, other participants see security as an \textit{implied requirement} that is always present. \textquote[PS04]{When the requirement is out but [privacy] has to be taken care of at every single step here. If someone comes to me asking for something then I assume that I do security for all the requirements. Wherever applicable security should be}. 

Security was also sometimes seen as a problem or requirement that should be solved by a designated entity within their workflow. For some participants, this entity was the \textit{operating system} 
\textquote[PS05]{Android, it is responsible. Because Android restricts my way of developing an application. So it should provide sufficient security mechanism for me to rely on}, \textquote[PS15]{Mostly the OS is the one that should provide security}. Others considered that a \textit{security team} in the software development workflow should be responsible. \textquote[PS04]{There should be a security team. Which takes care of that. Just like any other team inside the company. Like UI, testing team}. 

Many interviewees thought that the \textit{company} as a legal entity is responsible for S\&P, and some highlighted the role of legislation and government. \textquote[P17]{We are [responsible]. Not me personally but the company that I work for as a legal entity}. Moreover, a few saw \textit{end-users} having some responsibility as part of the larger S\&P ecosystem. \textquote[PS08]{There should be a certain amount of onus on the user, they should be responsible for like managing their password}. 

\subsubsection{General attitudes to APIs.}
Participants saw APIs as a useful and handy tool, especially in terms of code re-use \textquote[PS03]{GUI stuff in python, here you can just call functions without write whole part of code yourself. It's always 
handy}. APIs also allowed them to lean on the knowledge of others and not need to understand all the concepts themselves. \textquote[PS12]{It is quite useful and simple to import the library from platform. Before I used that library I need to learn each algorithm one by one mathematically. In terms of math the algorithm is quite hard. With library I just can, I import them from Internet. With one or two lines code I can use them. 
I can focus more on main procedure of neural network and data manipulation, so I can save a lot of time with the library}. 

Other peoples' code was a large theme when discussing APIs, particularly examples posted online or documentation-like guidance from others. \textquote[PS05]{Sometimes just some posts either forums or some question and answer community like Stack Overflow. There are people show you how to use in their answers, kind of you can copy paste and modify that to suit your needs}. APIs also tended to be designed in such a way that they were easy to start using. \textquote[PS01]{Maybe it is just experience, that makes it easier, because I was using APIs for so long so it is easy now to just come and start}. APIs also made it easy to get code running quickly, especially if the documentation was good and contained examples. \textquote[PS13]{If you pick a certain thing, you read the documentation, hopefully the documentation is done well, by done well I mean by examples. That you can get something to run as fast as possible because that keeps you motivated}.

\subsubsection{Security APIs.}
When asked about a `security API' participants struggled to understand what that could even be, falling back on areas commonly associated with security, like finance. \textquote[PS04]{What do you mean by security APIs? Something like payment gateway?}. Only one participant had a hands on experience with a security API which was problematic. \textquote[P19]{There is no feedback [in Android certificate validation]. It is a complete nightmare, various very long complicated classes archaic options that you are supposed to set. All and all was 40-50 lines of code. This was just a block of imperative commands for doing something basic like I'd like to validate against certificate file please. Absolutely crazy}. While only observed from one participant, his comments closely match what other researchers have observed from professional developers~\cite{egele2013empirical,arzt2015towards,fahl2013rethinking,georgiev2012most}.

P19, who has industry experience both as a developer and an intern, was one of the few people who discussed issues around secure programming, such as buffer overflow and functions with known security issues. \textquote[P19]{buffer overflows, system calls are an issue of languages, actually more that anything else. We still use C this is an atrocity, we shouldn't be using C anymore}. He is referring to common C function calls like \texttt{gets} which are impossible to use in a secure way, but are still commonly used due to being part of core C~\cite{kernighan2006}.

\subsubsection{Trusting other peoples' code.}
Using APIs and examples from the Internet was convenient for our participants, but it also required them to trust people they had never met. Some were concerned about blindly trusting code from unknown sources, but many had no problem instead choosing to trust in collective intelligence. \textquote[PS08]{If I download, I am often downloading source codes myself from the Internet and then building it. And again I don't have the time or the skill to audit say a code base that has millions of lines. I perhaps trust a little bit too much the crowd of people. If I look at the code base and see something on Github and it has let's say 2000 stars. Few hundred people watching it. The code is all open. I tend to perhaps foolishly I assume that if this many people have looked at it and if there was something up. Surely someone would do have said something. Download the code and build it. So it is possible that I have exposed myself to security issues as a result of that}. PS08 is referring to the `many eyeballs' idea in open source software which is an indicator of security and reliability of code for some developers~\cite{hissam2002trust}.

Trust is an inherent component of open source, that is code is open for everyone to read. \textquote[PS05]{As one of the 
reason I really want an open source app to do this is that this kind of app is allowed to access a lot of info. I don't trust any closed source software. I could use them but I don't trust them. Open source is the only way I could trust software. Although open source you could still add malicious code to open source in hope that people wouldn't discover that. But this is the only way}.

Trust in open source reaches to its highest level when people prefer to write less code and reuse others' code instead. \textquote[PS15]{So my idea is that the least I code the better. As long as [hypothetical app] is still maintained and supported regularly and I do update that regularly. Then I think I will be fine. Because tools that are widely used are very exposed to criticism so their maintainers usually patch up and correct their mistakes as fast as they can. So I'd very aware of what of dangers of the whole thing. And I would be careful to following news. But I'd avoid writing my own code} is a comment on open source software while the participant was discussing her hypothetical classroom app.

\section{Discussion}
\label{sec:discussion}

\subsubsection{Security mindsets.} 
Mindsets are likely to influence actions and decision making~\cite{wash2010folk,marki2016increasing,maier17influence}. We found that most students did not have a clearly developed concept of security. In fact, some participants even struggled to come up with words that could be associated with the term `computer security'. When it comes to threats such as hackers, what they can do, their intentions and capabilities is another point which needs improvements, we observe the similar patterns and folk models in CS students that others have seen in home users~\cite{wash2010folk,mazurek2010access}.

Mental models could be partially rooted in media~\cite{fulton2019effect}; participants cited media plot elements when describing hackers. End user security has seen success in teaching users to copy existing mental models such as viruses or home safety to better understand and reason about security and privacy~\cite{camp2009}. Our results suggest that similar approaches may work in the educational context to improve the mental models of students. 

\subsubsection{APIs.} 
When it comes to APIs, our results closely mirror what related work has shown for professional developers. They often use a combination of online resources to learn and use APIs. They prefer to use easier to use resources, and because official 
documentation is often not easy to use they tend to go for online resources like Stack Overflow~\cite{acar2016impact}. Professional developers (like our student sample) prefer documentation with examples and matching API scenarios~\cite{robillard2011field,Patnaik2019Usability}. Therefore, API designers are a significant element in secure software development ecosystem particularly industry API designers who have a large impact on developers. By designing usable APIs~\cite{green2016developers} and easy to understand documentation~\cite{robillard2011field} they can help students and developers learn and use APIs correctly which could result in building secure software.

\subsubsection{Division of labour.}
Who is in charge of doing security at organisations has long been a problem point with different units often thinking that security is the job of another team~\cite{arce2014avoiding}. A view shared by several of our participants. Though such a tendency is considered to be a ``key inhibitor toward secure software development practices''~\cite[p.~164]{xie2011programmers}.

In the work place, security auditors are in charge of checking code for issues which developers are then in charge of fixing~\cite{thomas2018security}. 
However this system has some downsides. First, auditing takes time during which developers work on other projects and loose the working memory they had about particular code segments. And second, fixing the code requires an understanding of the security issue in order to properly address it, and as has been previously shown, developers have difficulty interacting with security technologies like cryptography libraries due to misunderstandings around how cryptography works~\cite{egele2013empirical}.

In industry~\cite{assal_think_2019} it is necessary to create a security culture where basic security is everyone's responsibility and the security team is a component of that culture rather than the only people who `do' security. In education such a culture might be facilitated by providing student with code samples that are secure by default and by having them use code checking tools in IDEs that check for problems, such as static analysis tools which teach them not only that they should look for these issues, but also how.

Companies with high security standards make security as a commitment, do not satisfy security because of complexity, and they follow strict formal development and testing processes~\cite{haney2018we}. Universities can benefit these best practices and tech CS students how to become developers that care about S\&P.

\subsubsection{Security as a requirement.}
There are several similarities between the students' views and general industry practices. Student developers' treatment of security as an implied requirement is in line with findings that 
security is often treated as a non-functional feature in agile methods~\cite{fitzgerald2017continuous,bell2017agile}, and that the requirement is not explicitly stated~\cite{bartsch2011practitioners,bowen2014formality}. When asked to describe the features of the classroom discussion app, which had been intentionally chosen as an example of a task with implicit S\&P requirements, many students did not consider S\&P as an initial priority. For some students, this might be an artefact of their classroom development experience, where they tend to work on well formed projects that are unlikely to have security as an explicit requirement. 

Poor and inconsistent understanding of S\&P among CS students is likely to cause conflicts between real and best practices in the software industry. For example, when choosing a framework developers do not consider security as a deciding factor which contradicts secure development best practices~\cite{assal2018security}. 
In alignment with other aspects of software development, there is a need to synchronise the development approaches taught in the classroom with those used by industry. That synchronisation needs to occur in both directions such that students are taught industry best standards which they are then able to apply. 

\subsubsection{Internships.}
Internships are a way to engage students in the topic as well as prepare them for future careers~\cite{binder2015academic,chillas2015learning}. Although they require investments from industry~\cite{hoffman2012holistically} we believe that the shortage of S\&P professionals~\cite{furnell2017can} cannot be solved without involving every player. Hence, we encourage industry to offer more internships to CS students in S\&P fields to improve the number of students graduating with that type of experience.

\section{Limitations}
\label{sec:limitations}
Our population includes only students at a single Russell Group university in the UK. Even though our sample was diverse, it was not balanced for gender or security experience. Moreover, only two of our participants were native speakers of English, and we might have obtained more finely differentiated views and opinions if we had been able to interview each participant in their native language. Since we conducted the study during summer vacation time, this resulted in a participant pool biased towards Masters and PhD students, since undergraduate students are not normally present at University in the summer months. Possibly some of our potential participants were in their hometown and could not take part in this study.

\section{Future work}
\label{futureWork}
We plan to expand our study to other universities with a large scale survey to investigate differences and similarities across curriculum, universities and countries. Extending outcomes of this research to industry and professional developers and comparing results is also a path that could lead to valuable insights. Another interesting avenue for future work is to investigate the impact of open source and code reuse in system security. It also remains to question how developers trust in others' code and import code from different resources without knowing their source and coder.

\section{Conclusions}
\label{conclusions}
In this work we reported on a qualitative analysis of twenty semi-structured interviews with CS students. We find that the attitudes of students match many of those observed by other researchers looking at professional level developers. Students have a range of hacker/attack mindsets, lack of experience with security APIs, a mixed view of who is in charge of S\&P in the software life cycle, and a tendency to trust other peoples' code as a convenient approach to rapidly build software. We further give recommendations for both industry and academia to improve software S\&P ecosystem.

\section*{Acknowledgements}
Thanks to all participants for their time and everyone associated with the TULiPS Lab at the University of Edinburgh for helpful discussions and feedback. We also thank the anonymous reviewers whose comments helped improve the paper greatly. This work was sponsored in part by Microsoft Research through its PhD Scholarship Programme.

\bibliographystyle{splncs04}
\bibliography{bibliography.bib}

\input{supplementary.tex}

\end{document}

%% file: supplementary.tex
\clearpage
\section*{Appendix: interview script}
\label{interview-guideline}

\begin{enumerate}

\item Background \\
$\bullet$ Can you tell me about yourself? Your academic and professional background? $\bullet$ Can you tell me about your dream job?
\item App scenario \\
Let's say you were asked to create a new group discussion app for in-class discussions. 
$\bullet$ Free list: what features would you consider in this app?
$\bullet$ Here is a red pen. Can you circle the features that are security and privacy related? Or where you might have to consider security and privacy when building them?
$\bullet$ Why these ones?
$\bullet$ Who is most likely to try and attack this system? What are they likely going to try and do? 
\item Threats and attacks \\
$\bullet$ Can you tell me who hackers are, in your opinion? 
$\bullet$ Their intentions?
$\bullet$ What are hackers trying to get?
$\bullet$ Their background?
\item Responsibility attribution \\
$\bullet$ Who is responsible for providing security and privacy to end users? 
\item Prior coding experiences \\
$\bullet$ Tell me about the last piece of software you wrote.
$\bullet$ Did you consider security while building your project? If not this one, any other projects? 
$\bullet$ Can you tell me an example of an API/library? Can you give me some experiences you have had with them? Any experience with security APIs in particular?
$\bullet$ What was good about it? Why did you like it?
$\bullet$ What was confusing about it?
\item Personal security/privacy practices \\
Now we are going to switch to talking about how you handle security and privacy personally as an end user.  
$\bullet$ Free list: What words and concepts do you associate with computer security?
$\bullet$ Can you give me an example of a good computer security practice? What about something you have done yourself?
$\bullet$ Have you ever experienced a security or privacy compromise such as getting a virus on your computer, losing your password, having an email sent from your account, or loss of data about you? 
$\bullet$ How did you find out about the issue? 
$\bullet$ How did you correct it?
$\bullet$ What did you learn from the experience? 
$\bullet$ Can you tell me some about the experiences you have had with passwords?
\item Background and demographics \\
$\bullet$ How old are you?
$\bullet$ What is your degree title?
$\bullet$ Which year of the program are you in?
$\bullet$ What programming languages do you know?
$\bullet$ What programming courses have you taken?
$\bullet$ What security courses have you taken?
$\bullet$ What is your nationality? 
$\bullet$ Where did you study your undergraduate, Masters, or other degrees?

\end{enumerate}


%% file: main.bbl
\begin{thebibliography}{10}
\providecommand{\url}[1]{\texttt{#1}}
\providecommand{\urlprefix}{URL }
\providecommand{\doi}[1]{https://doi.org/#1}

\bibitem{acar2017comparing}
Acar, Y., Backes, M., Fahl, S., Garfinkel, S., Kim, D., Mazurek, M.L.,
  Stransky, C.: Comparing the {Usability} of {Cryptographic} {APIs}. In: {IEEE}
  {Symposium} on {Security} and {Privacy}. pp. 154--171 (2017)

\bibitem{acar2016impact}
Acar, Y., Backes, M., Fahl, S., Kim, D., Mazurek, M.L., Stransky, C.: You {Get}
  {Where} {You}'re {Looking} for: {The} {Impact} of {Information} {Sources} on
  {Code} {Security}. In: {2016 IEEE Symposium on Security and Privacy (SP)}.
  pp. 289--305 (2016)

\bibitem{acar2016dev}
Acar, Y., Fahl, S., Mazurek, M.L.: {You are Not Your Developer, Either: A
  Research Agenda for Usable Security and Privacy Research Beyond End Users}.
  In: {Cybersecurity Development (SecDev), IEEE}. pp.~3--8 (2016)

\bibitem{acar2017developers}
Acar, Y., Stransky, C., Wermke, D., Weir, C., Mazurek, M.L., Fahl, S.:
  Developers {Need} {Support}, {Too}: {A} {Survey} of {Security} {Advice} for
  {Software} {Developers}. In: {Cybersecurity Development (SecDev), 2017 IEEE}.
  pp. 22--26 (2017)

\bibitem{arce2014avoiding}
Arce, I., Clark-Fisher, K., Daswani, N., DelGrosso, J., Dhillon, D., Kern, C.,
  Kohno, T., Landwehr, C., McGraw, G., Schoenfield, B., et~al.: {Avoiding the
  Top 10 Software Security Design Flaws}. Technical report, IEEE Computer
  Societys Center for Secure Design (CSD)  (2014)

\bibitem{arzt2015towards}
Arzt, S., Nadi, S., Ali, K., Bodden, E., Erdweg, S., Mezini, M.: {Towards
  Secure Integration of Cryptographic Software}. In: 2015 ACM International
  Symposium on New Ideas, New Paradigms, and Reflections on Programming and
  Software (Onward!). pp. 1--13 (2015)

\bibitem{asgharpour2007mental}
Asgharpour, F., Liu, D., Camp, L.J.: {Mental Models of Security Risks}. In:
  Dietrich, S., Dhamija, R. (eds.) Financial Cryptography and Data Security.
  pp. 367--377. Springer Berlin Heidelberg (2007)

\bibitem{assal2018security}
Assal, H., Chiasson, S.: {Security in the Software Development Lifecycle}. In:
  Fourteenth Symposium on Usable Privacy and Security ({SOUPS}) (2018)

\bibitem{assal_think_2019}
Assal, H., Chiasson, S.: `{Think} secure from the beginning': {A} {Survey} with
  {Software} {Developers}. In: Proceedings of the 2019 {CHI} {Conference} on
  {Human} {Factors} in {Computing} {Systems} (2019)

\bibitem{Azhar2019Securing}
Azhar, M., Bhatia, S., Gagne, G., Kari, C., Maguire, J., Mountrouidou, X.,
  Tudor, L., Vosen, D., Yuen, T.T.: Securing the human: Broadening diversity in
  cybersecurity. In: Proceedings of the 2019 ACM Conference on Innovation and
  Technology in Computer Science Education. pp. 251--252 (2019)

\bibitem{balebako2014improving}
Balebako, R., Cranor, L.: {Improving App Privacy: Nudging App Developers to
  Protect User Privacy}. IEEE Security Privacy  12(4),  55--58 (2014)

\bibitem{bartsch2011practitioners}
Bartsch, S.: {Practitioners' Perspectives on Security in Agile Development}.
  In: Proceedings of the 2011 Sixth International Conference on Availability,
  Reliability and Security. pp. 479--484 (2011)

\bibitem{bell2017agile}
Bell, L., Brunton-Spall, M., Smith, R., Bird, J.: {Agile Application Security:
  Enabling Security in a Continuous Delivery Pipeline}. O'Reilly Media (2017)

\bibitem{binder2015academic}
Binder, J.F., Baguley, T., Crook, C., Miller, F.: The academic value of
  internships: Benefits across disciplines and student backgrounds.
  Contemporary Educational Psychology  41,  73--82 (2015)

\bibitem{bowen2014formality}
Bowen, J.P., Hinchey, M., Janicke, H., Ward, M.P., Zedan, H.: {Formality,
  Agility, Security, and Evolution in Software Development}. IEEE Computer
  47(10) (2014)

\bibitem{cambazoglu2013computer}
Cambazoglu, V., Thota, N.: {Computer Science Students' Perception of Computer
  Network Security}. In: Learning and Teaching in Computing and Engineering
  (LaTiCE). pp. 204--207. IEEE (2013)

\bibitem{camp2009}
{Camp}, L.J.: {Mental models of privacy and security}. IEEE Technology and
  Society Magazine  28(3),  37--46 (2009)

\bibitem{chillas2015learning}
Chillas, S., Marks, A., Galloway, L.: {Learning to labour: an evaluation of
  internships and employability in the ICT sector}. New Technology, Work and
  Employment  30(1),  1--15 (2015)

\bibitem{egele2013empirical}
Egele, M., Brumley, D., Fratantonio, Y., Kruegel, C.: {An Empirical Study of
  Cryptographic Misuse in Android Applications}. In: Proceedings of the 2013
  ACM SIGSAC Conference on Computer and Communications Security. pp. 73--84
  (2013)

\bibitem{gdpr2018eu}
The European parliament and the council of the European union: General Data
  Protection Regulation (GDPR) (2018),
  \url{https://eur-lex.europa.eu/legal-content/EN/TXT/PDF/?uri=CELEX:32016R0679}
  (Accessed August 2019)

\bibitem{fahl2012eve}
Fahl, S., Harbach, M., Muders, T., Baumg{\"a}rtner, L., Freisleben, B., Smith,
  M.: {Why Eve and Mallory love Android: An analysis of Android SSL (in)
  security}. In: Proceedings of the 2012 ACM conference on Computer and
  communications security. pp. 50--61 (2012)

\bibitem{fahl2013rethinking}
Fahl, S., Harbach, M., Perl, H., Koetter, M., Smith, M.: { Rethinking SSL
  Development in an Appified World}. In: Proceedings of the 2013 ACM SIGSAC
  conference on Computer \& Communications security. pp. 49--60 (2013)

\bibitem{fischer2017stack}
Fischer, F., B{\"o}ttinger, K., Xiao, H., Stransky, C., Acar, Y., Backes, M.,
  Fahl, S.: {Stack Overflow Considered Harmful? The Impact of Copy Paste on
  Android Application Security}. In: 2017 IEEE Symposium on Security and
  Privacy (SP). pp. 121--136 (2017)

\bibitem{fitzgerald2017continuous}
Fitzgerald, B., Stol, K.J.: Continuous software engineering: A roadmap and
  agenda. Journal of Systems and Software  123,  176--189 (2017)

\bibitem{fulton2019effect}
Fulton, K.R., Gelles, R., McKay, A., Abdi, Y., Roberts, R., Mazurek, M.L.: {The
  Effect of Entertainment Media on Mental Models of Computer Security}. In:
  {Fifteenth Symposium on Usable Privacy and Security ({SOUPS})} (2019)

\bibitem{furnell2017can}
Furnell, S., Fischer, P., Finch, A.: Can't get the staff? the growing need for
  cyber-security skills. Computer Fraud \& Security  2017(2),  5--10 (2017)

\bibitem{georgiev2012most}
Georgiev, M., Iyengar, S., Jana, S., Anubhai, R., Boneh, D., Shmatikov, V.:
  {The Most Dangerous Code in the World: Validating SSL Certificates in
  Non-browser Software}. In: Proceedings of the 2012 ACM Conference on Computer
  and Communications Security. pp. 38--49 (2012)

\bibitem{google2018diversity}
Google: Google diversity annual report (2018),
  \url{http://diversity.google/annual-report} (Accessed August 2019)

\bibitem{green2016developers}
Green, M., Smith, M.: Developers are {Not} the {Enemy}!: {The} {Need} for
  {Usable} {Security} {APIs}. IEEE Security \& Privacy  14(5),  40--46 (2016)

\bibitem{hadar2018privacy}
Hadar, I., Hasson, T., Ayalon, O., Toch, E., Birnhack, M., Sherman, S.,
  Balissa, A.: Privacy by designers: software developers' privacy mindset.
  Empirical Software Engineering  23(1),  259--289 (2018)

\bibitem{haney2018we}
Haney, J.M., Theofanos, M., Acar, Y., Prettyman, S.S.: {``We make it a big deal
  in the company'': Security Mindsets in Organizations that Develop
  Cryptographic Products}. In: Fourteenth Symposium on Usable Privacy and
  Security ({SOUPS}) (2018)

\bibitem{hissam2002trust}
Hissam, S.A., Plakosh, D., Weinstock, C.: Trust and vulnerability in open
  source software. IEE Proceedings-Software  149(1),  47--51 (2002)

\bibitem{hoffman2012holistically}
Hoffman, L., Burley, D., Toregas, C.: {Holistically Building the Cybersecurity
  Workforce}. IEEE Security \& Privacy  10(2),  33--39 (2012)

\bibitem{iacono2017and}
Iacono, L.L., Gorski, P.L.: {I Do and I Understand. Not Yet True for Security
  APIs. So Sad}. In: Proc. of the 2nd European Workshop on Usable Security,
  ser. EuroUSEC (2017)

\bibitem{indela2016toward}
Indela, S., Kulkarni, M., Nayak, K., Dumitras, T.: {Toward Semantic
  Cryptography APIs}. In: Cybersecurity Development (SecDev), IEEE. pp. 9--14
  (2016)

\bibitem{icoDataAnalytics}
{Information Commissioner's Office}: Investigation into the use of data
  analytics in political campaigns (2018),
  \url{https://ico.org.uk/media/action-weve-taken/2259371/investigation-into-data-analytics-for-political-purposes-update.pdf}
  (Accessed August 2019)

\bibitem{jones2017should}
Jones, K., Siami~Namin, A., Armstrong, M.: {What Should Cybersecurity Students
  Learn in School?: Results from Interviews with Cyber Professionals}. In:
  Proceedings of the 2017 ACM SIGCSE Technical Symposium on Computer Science
  Education. pp. 711--711 (2017)

\bibitem{kernighan2006}
Kernighan, B.W., Ritchie, D.M.: {The C Programming Language}. Prentice Hall
  (2006)

\bibitem{lazar2014does}
Lazar, D., Chen, H., Wang, X., Zeldovich, N.: {Why Does Cryptographic Software
  Fail?: A Case Study and Open Problems}. In: Proceedings of 5th Asia-Pacific
  Workshop on Systems. p.~7. ACM (2014)

\bibitem{lazar2017research}
Lazar, J., Feng, J.H., Hochheiser, H.: {Research Methods in Human-Computer
  Interaction}. Morgan Kaufmann (2017)

\bibitem{maier17influence}
Maier, J., Padmos, A., Bargh, M.S., W\"orndl, W.: {Influence of Mental Models
  on the Design of Cyber Security Dashboards}. In: Proceedings of the 12th
  International Joint Conference on Computer Vision, Imaging and Computer
  Graphics Theory and Applications - Volume 3: IVAPP, (VISIGRAPP). pp. 128--139
  (2017)

\bibitem{mansfield2015ashley}
Mansfield-Devine, S.: {The Ashley Madison affair}. Network Security  2015(9),
  8--16 (2015)

\bibitem{marki2016increasing}
M{\"a}rki, H., Maas, M., Kauer-Franz, M., Oberle, M.: {Increasing Software
  Security by Using Mental Models}. In: Advances in Human Factors in
  Cybersecurity, pp. 347--359. Springer (2016)

\bibitem{mazurek2010access}
Mazurek, M.L., Arsenault, J.P., Bresee, J., Gupta, N., Ion, I., Johns, C., Lee,
  D., Liang, Y., Olsen, J., Salmon, B., Shay, R., Vaniea, K., Bauer, L.,
  Cranor, L.F., Ganger, G.R., Reiter, M.K.: {Access Control for Home Data
  Sharing: Attitudes, Needs and Practices}. In: Proceedings of the SIGCHI
  Conference on Human Factors in Computing Systems. pp. 645--654 (2010)

\bibitem{miles1994qualitative}
Miles, M., Huberman, M.: {Qualitative Data Analysis: A Methods Sourcebook}.
  Sage (1994)

\bibitem{munro2015kettles}
Munro, K.: {Hacking kettles \& extracting plain text WPA PSKs. Yes really!}
  (2015),
  \url{https://www.pentestpartners.com/security-blog/hacking-kettles-extracting-plain-text-wpa-psks-yes-really}
  (Accessed August 2019)

\bibitem{nadi2016jumping}
Nadi, S., Kr{\"u}ger, S., Mezini, M., Bodden, E.: Jumping {Through} {Hoops}:
  {Why} {Do} {Java} {Developers} {Struggle} with {Cryptography} {APIs}? In:
  Proceedings of the 38th International Conference on Software Engineering. pp.
  935--946. ACM (2016)

\bibitem{nielson2016playground}
Nielson, S.J.: {PLAYGROUND: preparing students for the cyber battleground}.
  Computer Science Education  26(4),  255--276 (2016)

\bibitem{oliveira2014s}
Oliveira, D., Rosenthal, M., Morin, N., Yeh, K.C., Cappos, J., Zhuang, Y.:
  {It's the Psychology Stupid: How Heuristics Explain Software Vulnerabilities
  and How Priming Can Illuminate Developer's Blind Spots}. In: Proceedings of
  the 30th Annual Computer Security Applications Conference. pp. 296--305
  (2014)

\bibitem{oltrogge2018rise}
Oltrogge, M., Derr, E., Stransky, C., Acar, Y., Fahl, S., Rossow, C.,
  Pellegrino, G., Bugiel, S., Backes, M.: {The Rise of the Citizen Developer:
  Assessing the Security Impact of Online App Generators}. In: 2018 IEEE
  Symposium on Security and Privacy (SP). pp. 634--647 (2018)

\bibitem{owasp2017top}
OWASP: {Top 10 Most Critical Web Application Security Risks}. Tech. rep., {{The
  OWASP Foundation}} (2017)

\bibitem{Patnaik2019Usability}
Patnaik, N., Hallett, J., Rashid, A.: {Usability Smells: An Analysis of
  Developers{\textquoteright} Struggle With Crypto Libraries}. In: Fifteenth
  Symposium on Usable Privacy and Security ({SOUPS}) (2019)

\bibitem{pieczul2017developer}
Pieczul, O., Foley, S., Zurko, M.E.: {Developer-centered Security and the
  Symmetry of Ignorance}. In: Proceedings of the 2017 New Security Paradigms
  Workshop. pp. 46--56. ACM (2017)

\bibitem{poller2017can}
Poller, A., Kocksch, L., T{\"u}rpe, S., Epp, F.A., Kinder-Kurlanda, K.: {Can
  Security Become a Routine?: A Study of Organizational Change in an Agile
  Software Development Group}. In: Proceedings of the 2017 ACM Conference on
  Computer Supported Cooperative Work and Social Computing. pp. 2489--2503
  (2017)

\bibitem{radermacher2013gaps}
Radermacher, A., Walia, G.: {Gaps Between Industry Expectations and the
  Abilities of Graduates}. In: Proceeding of the 44th ACM Technical Symposium
  on Computer Science Education. pp. 525--530 (2013)

\bibitem{radermacher2014investigating}
Radermacher, A., Walia, G., Knudson, D.: {Investigating the Skill Gap Between
  Graduating Students and Industry Expectations}. In: Companion Proceedings of
  the 36th International Conference on Software Engineering. pp. 291--300. ACM
  (2014)

\bibitem{renaud2014doesn}
Renaud, K., Volkamer, M., Renkema-Padmos, A.: {Why Doesn’t Jane Protect Her
  Privacy?} In: International Symposium on Privacy Enhancing Technologies
  Symposium. pp. 244--262. Springer (2014)

\bibitem{robillard2011field}
Robillard, M.P., Deline, R.: {A field study of API learning obstacles}.
  Empirical Software Engineering  16(6),  703--732 (2011)

\bibitem{saldana2015coding}
Salda{\~n}a, J.: The Coding Manual for Qualitative Researchers. Sage (2015)

\bibitem{stackoverflow2019dev}
StackOverflow: Developer Survey Results (2019),
  \url{https://insights.stackoverflow.com/survey/2019} (Accessed August 2019)

\bibitem{sudol2010analyzing}
Sudol, L.A., Jaspan, C.: {Analyzing the Strength of Undergraduate
  Misconceptions About Software Engineering}. In: Proceedings of the Sixth
  International Workshop on Computing Education Research. pp. 31--40 (2010)

\bibitem{tabassum2018evaluating}
Tabassum, M., Watson, S., Chu, B., Lipford, H.R.: {Evaluating Two Methods for
  Integrating Secure Programming Education}. In: Proceedings of the 49th ACM
  Technical Symposium on Computer Science Education. pp. 390--395 (2018)

\bibitem{thomas2018security}
Thomas, T.W., Tabassum, M., Chu, B., Lipford, H.: {Security During Application
  Development: an Application Security Expert Perspective}. In: Proceedings of
  the 2018 CHI Conference on Human Factors in Computing Systems. p.~262 (2018)

\bibitem{ukrop2018johnny}
Ukrop, M., Matyas, V.: {Why Johnny the Developer Can't Work with Public Key
  Certificates}. In: Cryptographers' Track at the RSA Conference. Springer
  (2018)

\bibitem{vaniea_betrayed_2014}
Vaniea, K.E., Rader, E., Wash, R.: Betrayed by {Updates}: {How} {Negative}
  {Experiences} {Affect} {Future} {Security}. In: Proceedings of the {SIGCHI}
  {Conference} on {Human} {Factors} in {Computing} {Systems} (2014)

\bibitem{vox2018facebook}
Vox: {The Cambridge Analytica Facebook scandal} (2018),
  \url{https://www.vox.com/2018/4/10/17207394} (Accessed August 2019)

\bibitem{wash2010folk}
Wash, R.: {Folk Models of Home Computer Security}. In: Proceedings of the Sixth
  Symposium on Usable Privacy and Security (SOUPS) (2010)

\bibitem{weir2016improve}
Weir, C., Rashid, A., Noble, J.: {How to Improve the Security Skills of Mobile
  App Developers: Comparing and Contrasting Expert Views}. In: Twelfth
  Symposium on Usable Privacy and Security ({SOUPS}) (2016)

\bibitem{whitney2015embedding}
Whitney, M., Lipford-Richter, H., Chu, B., Zhu, J.: {Embedding Secure Coding
  Instruction into the IDE: A Field Study in an Advanced CS Course}. In:
  Proceedings of the 46th ACM Technical Symposium on Computer Science Education
  (2015)

\bibitem{wurster2009developer}
Wurster, G., van Oorschot, P.C.: The {Developer} is the {Enemy}. In:
  Proceedings of the 2008 New Security Paradigms Workshop. pp. 89--97. ACM
  (2009)

\bibitem{xie2011programmers}
Xie, J., Lipford, H.R., Chu, B.: Why do programmers make security errors? In:
  2011 IEEE Symposium on Visual Languages and Human-Centric Computing (VL/HCC).
  pp. 161--164 (2011)

\bibitem{zou2018equifax}
Zou, Y., Mhaidli, A.H., McCall, A., Schaub, F.: {``I've Got Nothing to Lose'':
  Consumers Risk Perceptions and Protective Actions after the Equifax Data
  Breach}. In: Fourteenth Symposium on Usable Privacy and Security ({SOUPS})
  (2018)

\end{thebibliography}
